\documentclass[numbers,floatfix,nopreprint]{jasatex}
\usepackage{amsmath,amsfonts,epsfig,bm}
\usepackage[latin1]{inputenc}

\newcommand{\argmin}{\mathop{\mathrm{argmin}}\limits}
\newcommand{\st}{\quad \mbox{subject to} \quad}

\begin{document}

\title{Nearfield Acoustic Holography using sparsity and compressive sampling principles}
\author{Gilles Chardon}
\affiliation{Institut Langevin, ESPCI ParisTech - Univ Paris Diderot - UPMC Univ Paris 06 - CNRS UMR 7587, 10 rue Vauquelin F-75005 Paris France}
\email{gilles.chardon@espci.fr}
\author{Laurent Daudet}
\affiliation{Institut Langevin, ESPCI ParisTech - Univ Paris Diderot - UPMC Univ Paris 06 - CNRS UMR 7587, 10 rue Vauquelin F-75005 Paris France}
\author{Antoine Peillot}
\affiliation{UPMC Univ Paris 06, UMR 7190 - Institut Jean Le Rond d'Alembert, F-75005 Paris France.}
\author{François Ollivier}
\affiliation{UPMC Univ Paris 06, UMR 7190 - Institut Jean Le Rond d'Alembert, F-75005 Paris France.}
\author{Nancy Bertin}
\affiliation{CNRS - IRISA-UMR 6074, Campus de Beaulieu, F-35042 Rennes Cedex France}
\author{R\'emi Gribonval}
\affiliation{INRIA, Centre Inria Rennes - Bretagne Atlantique, Campus de Beaulieu, F-35042 Rennes Cedex France}

\begin{abstract}
Regularization of the inverse problem is a complex issue when using Near-field Acoustic Holography (NAH) techniques to identify the vibrating sources. This paper shows that, for convex homogeneous plates with arbitrary boundary conditions, new regularization schemes can be developed, based on the sparsity of the normal velocity of the plate in a well-designed basis, \emph{i.e.} the possibility to approximate it as a weighted sum of few elementary basis functions. In particular, these new techniques can handle discontinuities of the velocity field at the boundaries, which can be problematic with standard techniques. This comes at the cost of a higher computational complexity to solve the associated optimization problem, though it remains easily tractable with out-of-the-box software. Furthermore, this sparsity framework allows us to take advantage of the concept of Compressive Sampling: under some conditions on the sampling process (here, the design of a random array, which can be numerically and experimentally validated), it is possible to reconstruct the sparse signals with significantly less measurements (\emph{i.e.}, microphones) than classically required. 
  After introducing the different concepts, this paper presents numerical and experimental results of NAH with two plate geometries, and compares the advantages and limitations of these sparsity-based techniques over standard Tikhonov regularization.  
\end{abstract}

\maketitle

%% END_HEADER %%

{\renewcommand{\thefootnote}{}\footnotetext{\em The following article has been accepted by JASA. After it is published, it will be found at http://scitation.aip.org/JASA}}

\section{INTRODUCTION}
Nearfield Acoustic Holography (NAH), first introduced by Maynard \textit{et al}\cite{NAH1}, is a widespread method to measure the normal velocity of  vibrating structures, based on some 
measurements of the radiated soundfield at a close distance.
This set of pressure measurements, called hologram, is processed to fulfill two different goals.
At a given angular frequency $\omega$, one seeks numerically either the prediction of the far field by solving a direct propagation problem, or the reconstruction of the normal velocity distribution $\dot w$ of the source vibrating structure by solving an inverse back-propagation problem. While the former option presents no computational difficulty (it will not be discussed here), the latter one is subject to ill-conditioning and requires a regularization procedure. 

Fig. \ref{SetUp} shows the block diagram of NAH data acquisition and processing as discussed in this paper, as well as the block diagram of the control 
measurements using a laser vibrometer. Further details are given in section~\ref{sec:stdNAH}.

Although conceptually simple, at least in the case of a planar vibrating structure\cite{NAH1}, a practical use of NAH still requires a careful experimental design, some precise measurements and a non-trivial data processing generally carried out in the spatial frequency domain (also called $k$-space or wavenumber domain). In particular, the two following issues are commonly raised. As for most inverse problems, regularization is an essential component. In order to avoid the amplification of noise in evanescent waves, many regularization principles have been proposed, and this issue is still  a very active field of research \cite{ WilliamsRegul, KimJSV04, Scholte}. However, a usual side effect of regularization is the loss of high spatial frequencies, which damages the reconstruction of free ends for example.  
  Furthermore, to be able to image the vibration field at high frequencies, it is often found necessary to sample  finely the pressure field in the measurement plane. This may require a large number of microphones, with the associated issues of synchronization, calibration, A/D conversion, and total data throughput.

The goal of this paper is to demonstrate
that, in the case of a star-shaped homogeneous plate, sparse regularization principles and/or compressive sampling techniques lead to significant improvements for these two issues over standard NAH techniques. 
In fact, this processing can be viewed as an example of a model-based inverse problem, that have been developed in a large number of other contexts, for instance in geophysics \cite{Zwartjes}. In acoustics, the HELS method 
\cite{ZhaoJSV05} can also be seen as a model-based inversion, based on least squares. The key point of sparsity-based methods, which to the best of our knowledge have not been used in the context of NAH, is that they address the two issues of regularization and 
number of measurements within a unified theoretical framework. These methods have already been applied to a large variety of inverse problems,
 such as the one pixel camera \cite{singlepixel}, magnetic resonance imaging \cite{lustig08compressed} or synthetic aperture radar \cite{journals/jstsp/PatelEHC10}.

The downside of the proposed method over more generic methods is that the domain under study must be planar, star-shaped (which includes all convex shapes), and homogeneous.  
However, one should emphasize that it holds for any type of boundary conditions, that do not have to be known. For instance,  it can be used to study vibrating 
plates that are part of complex structures, regardless on how they are attached to the structure. 
Possible extensions to more complex geometries are also discussed. 

\medskip

 To summarize the outline of the paper, the three most important contributions can be stated as follows:
\begin{itemize}
  \item {\em Sparse regularization for NAH:} in a well-designed basis (called {\em dictionary}), the sought-after velocity field is approximately sparse, which means that it can be well approximated by a linear combination of a small number of elementary basis functions. The solutions obtained using this sparsity principle as a regularizer have, as we shall see,  a good fit with the reference data (laser velocimetry), without loss of spatial high frequencies. 
  We discuss in section~ \ref{se:sparsity} the design of a dictionary of elementary basis functions adapted to the considered problem.
  \item {\em Sub-Nyquist random sampling for NAH: proof of concept.} Within the sparse regularization framework, a fine sampling of the hologram plane is indeed not necessary. The theory of compressive sampling (CS, also called compressed sensing), presented in section~\ref{se:csnah}, gives a theoretical foundation on how to include sparsity priors directly at the signal acquisition stage. Thus, under the sparsity assumption, it is possible to significantly reduce the number of measurements (\emph{i.e.}, microphones), even well below spatial  Nyquist rates. However, CS theory tells us that, for a successful identification of the sparse components,  measurements have to be as incoherent as possible with the sparsity basis (in other words, each measurement must carry information about all significant coefficients). To satisfy this in our experimental setup, we first develop a proof of concept: we simulate measurements with a random array by randomly selecting a subset of microphones from a large regular antenna. The corresponding experimental results are described in section~\ref{se:results}, demonstrating the proposed CS approach to NAH, which combines a random subsampling of the hologram measurements with sparse regularization for the reconstruction of the velocity field.  
    \item {\em Experimental validation and discussion.} 
Finally, we design and build a new NAH array where a limited number of microphones are actually randomly placed in the plane above the vibrating structure.  The design of this array is described in section~\ref{se:randomarray}, with supporting numerical simulations and experimental results comparing the new antenna with the randomly subsampled design used in the proof of concept. Section~\ref{se:cafeducommerce} discusses the benefits of this approach for a precise identification of a vibration pattern (robustness to calibration errors, ability to reconstruct high spatial frequencies and discontinuities at the boundaries, reduced number of microphones, ...),  the new issues it raises (choice of an appropriate dictionary of basis functions, tuning of the parameters of the sparse regularization algorithms, numerical complexity of the algorithms, ...), as well as possible extensions to overcome some of theses difficulties and apply it in a wider context. 
\end{itemize}

\section{Standard implementations of NAH}
\label{sec:stdNAH}

In this section we describe the standard techniques for NAH, including Tikhonov regularization that will be used as baseline for comparison with the proposed sparse regularization method. 

\subsection{Measurement techniques}

A first matter to consider in collecting NAH data is whether the vibrating source is controlled or not, which sets the spectral support of the acoustic field to be processed. Some observe the structure in operational conditions, others apply controlled excitation, either harmonic to study the behavior at a given frequency, or with a wide frequency band support (impulsive or random) to collect a response over the same support, and to allow an exhaustive study such as structural modal analysis. 

Another issue regarding the measurement technique concerns the sensors to be used. While the sole mean for measuring the acoustic field has long been using microphones or arrays of microphones, some authors have recently discussed the advantages of using particle velocity sensors\cite{ZhangJacobsen, JacobsenJASA05, JacobsenJASA06}. 
Whatever choice is made for the sensors and type of excitation, the acquired signals must undergo a time and frequency analysis prior to the NAH process itself, in order to separate the harmonic acoustic fields of interest. 

\subsection{Geometries involved}

 When considering planar structures, the mathematical  formulation of NAH and its numerical implementation are simpler  and take advantage of very fast Discrete Fourier Transform (DFT) algorithms. Provided some adaptations are made, cylindrical and spherical  geometries can also be processed rather simply. These basic implementations are exhaustively described by Williams\cite{WilliamsBook}.
 
For objects with arbitrary shapes, it is necessary to carry out the NAH process using more complex methods, that involve large matrix inversions using singular value decomposition (SVD). This comes at the price of a high computational cost, \emph{e.g.} the inverse boundary element method (IBEM) \cite{SchuhJASA03, ValdiviaJASA06} based on the integral formulation of the radiation theory or the HELS method\cite{ZhaoJSV05}. 

\subsection{Mathematical formulation}

\begin{figure}[htbp]
\includegraphics[width = 8cm]{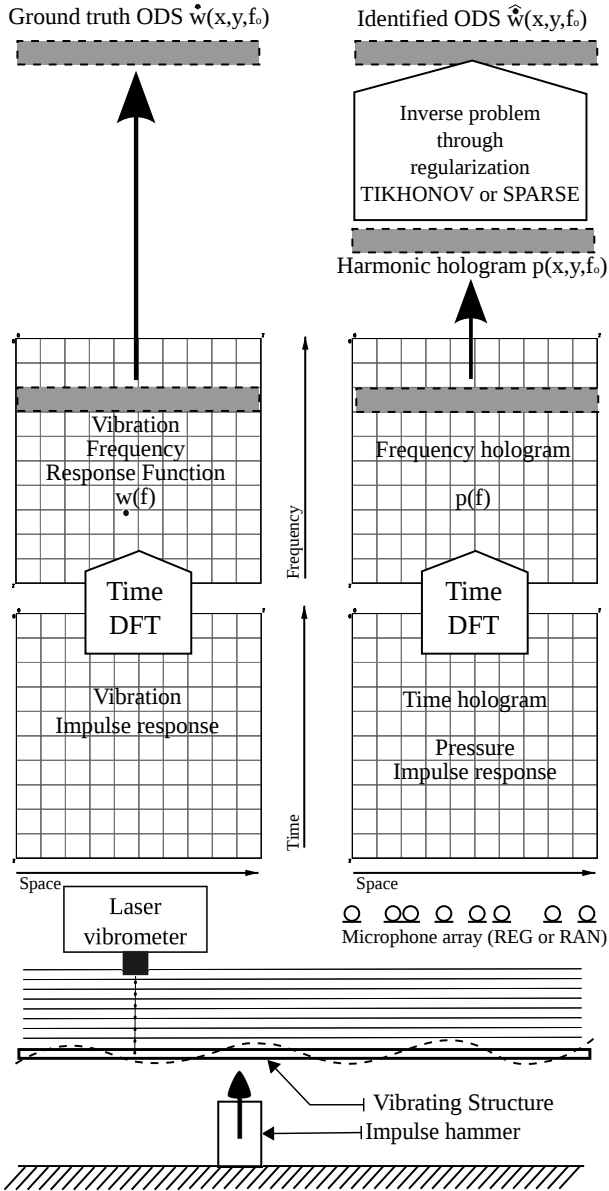}
\caption{Block diagram of the experimental process. The left column is the ``ground-truth'' reference measurement with laser vibrometer, the right column is the NAH processing.}
\label{SetUp}
\end{figure}

Pressure on a plane of elevation $z_0$ radiated by a  plate with a
distribution of normal velocities $\dot w (x,y, 0)$ of elevation $0$ at angular frequency $\omega$ is given by the convolution of the source
distribution with the propagator $g(x,y,z) = g(\vec r) = -i\rho c k\frac{e^{i k ||\vec r||}}{2\pi ||\vec r||}$  :

\begin{equation}
p(x,y,z_0) = g(x,y,z_0) \star_{xy} \dot w(x,y,0)
\label{prop}
\end{equation}
Here $\star_{xy}$ denotes 2-D convolution in the $x$ and $y$ variables,
$c$ the wave velocity, $\rho$ the air density and $k = \omega/c$ the wavenumber.  
The 2-D Spatial Fourier transform of this equation, with respect to  $x$ and $y$, yields :

\begin{equation}
P(k_x,k_y,z_0) = G(k_x, k_y, z_0)  \dot W(k_x, k_y, 0).
\label{propf}
\end{equation}
where $k_x$ an $k_y$ are the wavenumbers in the $x$ and $y$ directions
respectively.

The implementation of standard NAH uses a discretized formulation.
The pressure and plate velocity are sampled in a rectangular domain
(i.e., the antenna),
and their spatial Fourier transforms are approximated by the discrete
Fourier transforms of their sampled versions:
\begin{equation}
\mathbf{Fp = GF\dot w}
\end{equation}
where:
\begin{itemize}
\item $\mathbf {\dot w}$ denotes the vector of source normal velocities to be identified, discretized on a rectangular regular grid,
\item $\mathbf p$ is the vector of measured pressures, also discretized in the hologram plane,
\item $\mathbf F$ is the 2-D spatial DFT operator, and
\item $\mathbf G$ is zero except on the diagonal where it is equal to $G$ sampled at the wave vectors of the DFT basis vectors.
\end{itemize}
The pressure in function of the plate velocity writes
\begin{equation}
\mathbf{p = F^{-1}GF\dot w.}
\label{nah_eq}
\end{equation}
The matrix product $\mathbf{F^{-1}GF}$ will be noted $\mathbf{H}$ and its conjugate transpose $\mathbf{H}^\star$.
The resolution of the inverse problem provides an estimate of the normal velocity $\mathbf{\hat{\dot w}}$ of the structure. Naive inversion of Equation
\eqref{nah_eq} yields
\begin{equation}
 \mathbf {\hat{\dot w} = F^{-1}G^{-1}F p = H^{-1}p} 
\label{nah_eq2}
\end{equation}
where $\mathbf{G}$, being diagonal, is easily invertible. However, due to
its ill-conditioning, the computation of the sources using this equation
is very unstable, and thus requires regularization as described in the next section.

\subsection{Need for regularization} 
\label{sse:StdRegul}

The basic theory of NAH asks for the hologram to extend on a larger area than the source, so as to enclose the limits of the acoustic field produced by the vibrating source. In some cases, this constraint cannot be fulfilled and some precautions must be taken. Some authors study the reconstruction of the source limited to a region of interest from the measurement of a patch smaller than the source \cite{SarkisJASA05, ThomasPascal}. In this case, one has to take care of the leakage artifacts generated by the truncation of the field. This can be performed in various ways, the simplest being  to taper  the measured field with a flattop window (or Tukey window). Some use a wavelet-based pre-processing of the measured field to lessen the truncation artifacts 
\cite{ThomasPascal}. 

The hologram $p$ must be recorded at a short distance $z_0$, in order to collect the decisive information carried by evanescent waves~\cite{Bertero:1981aa}, which decay rapidly, \emph{e.g.} exponentially in the case of planar geometries. Yet, the measured field is generally contaminated by noise of various nature. 
Hence, the naive back-propagation process (Equation \eqref{nah_eq2}) ruins the reconstruction, as it implies the amplification of the evanescent components together with part of the measurement noise lying in the same $k$-region (high wavenumbers). It is therefore necessary to 
use a regularization step along with the NAH process. The simplest candidate for this task is a low-pass filter in the spectral domain. The optimization of such a filter has generated an abundant literature \cite{WilliamsRegul,KimJSV04, Scholte}. 
It essentially consists  in finding the cutoff frequency and slope of the filter, through the determination of a scalar regularization parameter. For this purpose, various algorithms are in competition; here, we briefly describe the most popular, called Tikhonov regularization, as implemented in the following experiments.

\subsection{Tikhonov regularization}
Tikhonov regularization, widely used in many ill-posed inverse problems, adds a penalty term to the inverse problem and, generally,
involves the resolution of the following minimization problem:
\begin{equation}
\label{eqn:tikhonovregul}
\hat{\dot{\mathbf{w}}} = 
\argmin_{ \dot{\mathbf{w}}} {\left\|\mathbf{p-H\dot{w}}\right\|}^2_{2} + \lambda{\left\|\mathbf{L}\mathbf{\dot{w}}\right\|}^2_{2}
\end{equation}
where $\mathbf{L}$ is the so-called Tikhonov matrix and $\lambda$ the regularization parameter. The result of the Tikhonov regularization can be expressed in closed form as
\begin{equation}
\label{eqn:tikhonovexplicit}
\hat{\dot{\mathbf{w}}} = \mathbf{R}_\lambda \mathbf{H}^{-1} \mathbf{p}
\end{equation}
where $\mathbf{R}_\lambda$ writes
\begin{equation}
\label{eqn:rl}
\mathbf{R}_\lambda = (\mathbf{H^\star H}+\lambda \mathbf{L}^\star\mathbf{L})^{-1}\mathbf{H^\star H}.
\end{equation}
$\mathbf H$ and its inverse cancel when equations \eqref{eqn:tikhonovexplicit} and \eqref{eqn:rl} are combined, avoiding
the computation of this ill-conditionned inverse. However, this expression of $\mathbf{R}_\lambda$ highlights the fact that,
in the basic application of the Tikhonov regularization to NAH where $\mathbf{L}$ is chosen to be the identity matrix,  $\mathbf{R}_\lambda$ acts as a low-pass spatial filter. For our study, we use an improved Tikhonov approach where $\mathbf{L}$ is chosen so that this filter is sharpened up \cite{WilliamsRegul}, and the Generalized Cross-Validation (GCV)\cite{WilliamsRegul, Scholte} is used
to estimate the regularization parameter.

Though numerous studies have proved that this standard implementation of NAH provides good results, there are still theoretical and 
practical issues. Two of them can be emphasized. First, it does not capture well the discontinuities at the boundaries when free boundary conditions are used: indeed, the regularization acts as a low-pass filter and therefore smoothes all discontinuities.  Second, at high frequencies, this techniques requires the sampling of the hologram plane on a fine grid to satisfy the Nyquist criterion; this may require a very high number of microphones (and corresponding A/D converters and acquisition channels). 
This study will show that, by setting this inverse problem into a sparse regularization framework, both issues can be alleviated. 
In the next section, we recall the concept of sparsity, and in particular we emphasize that the new formulation of the regularized inverse problem is very similar to a Tikhonov approach, up to a change of basis and a different norm for the penalty term. 
  
\section{NAH in a sparsity framework}
\label{se:sparsity}

In this section, we first give a brief description of the main notions regarding sparsity (section \ref{subse:sparsity}), and the corresponding optimization algorithms (section \ref{subse:algorithms}). We then investigate the choice of a {\em dictionary} of basis functions where the signal of interest can be sparsely approximated (section \ref{subse:dico}). Finally, this allows us to re-cast the NAH inverse problem into a 
sparsity-promoting optimization problem, which is the first original contribution of this work.

\subsection{Sparsity}
\label{subse:sparsity}
Sparsity is the property of a given signal to be decomposed as a linear combination of a small number of pre-defined basis functions, called {\em atoms}. 
 It is used in numerous applications\cite{mallat09wavelet}, ranging from data compression to source separation, signal analysis etc. 
More precisely, we call {\em dictionary} $\cal{D}$ a set, that we here assume of finite size $M$, of atoms 
$\mathbf{d}_k \in \mathbb{R}^N$:
 ${\cal{D}} = \{\mathbf{d}_k\}_{k = 1 \dots M}$. The dictionary $\cal{D}$ can be a basis of $\mathbb{R}^N$ ($M=N$), or an overcomplete family spanning $\mathbb{R}^N$ ($M > N$).

A finite-size discrete signal $\mathbf{x} \in \mathbb{R}^N$ is sparse in  $\cal{D}$ if there exists a set of coefficients  $\alpha_j$ such that 
\begin{equation}
\mathbf{x} = \sum_{j \in J} \alpha_j \mathbf{d}_j
\label{eq:sparsedec}
\end{equation}
where $J$ is a subset of $\{1 \dots M \}$, of much smaller size: $\sharp J = \operatorname{Card} (J) \ll M$. 
This decomposition can be exact (exact sparsity), or approximate (compressibility).
However, from a given signal $\mathbf{x}$, there is in general no unique decomposition 
such as Equation~\eqref{eq:sparsedec} if $\cal{D}$ is overcomplete. In order to find the sparsest set of coefficients $\alpha$  that verifies Equation~\eqref{eq:sparsedec}, or that provides the best balance between data fidelity and sparsity, a large number of algorithms have been developed, and in particular the $\ell_1$ optimization techniques that will be described in the next section. 

\medskip

In the case of NAH, we will assume that the discretized version of the Fourier-domain velocity map of the source plane $\hat{\dot{w}}$ is approximately sparse in an appropriate basis:
\begin{equation}
\hat{\dot{\mathbf{w}}} \approx \mathbf{D}\bm\alpha
\end{equation}
where the vector $\alpha$ has only $\sharp J$ non-zero elements, and $\mathbf{D} \in \mathbb{R}^{N \times M} $ is the matrix whose columns are all the $\mathbf{d}_k$ in $\cal{D}$.
As we shall see, this assumption is only justified for an appropriate choice of dictionary $\cal{D}$. This is a crucial point if one wants to use sparse decompositions, as this really influences how sparse a given class of signals can be: the sparser the decomposition (the smaller $\sharp J$), the better the estimation. 

Provided that this holds, the NAH inverse problem can be re-cast as follows : for a given set of pressure measurements $\mathbf{p}$ (the hologram), find the sparsest set of coefficients $\alpha$ such that  $\mathbf{p} = \mathbf{H D} \bm\alpha$
(note that, here, the matrix $\mathbf{H}$ is obtained 
by direct discretization of Equation \eqref{prop}, i.e. quadratures of the integrals used to compute
the convolution, avoiding truncature artifacts
of the Fourier transform). More precisely, if one introduces the $\ell_0$ pseudo-norm of the vector $\bm\alpha = \left[\alpha_1 \ldots \alpha_M\right]^T$ as $\|\bm\alpha\|_0 = \sharp J$ (number of nonzero coefficients), the problem to be solved is 
\begin{equation}
\label{eqn:nphardpb}
\argmin_{\bm\alpha} \|\bm\alpha\|_0 \st  \quad \mathbf{p} = \mathbf{H D} \bm\alpha,
\end{equation}
Unfortunately, as the $\ell_0$ pseudo-norm is non convex, and in general exactly solving this problem would require a combinatorial exhaustive search, that immediately becomes non-tractable. In the next section, we describe the algorithm we used to provide practical solutions to this problem. 
 
\subsection{$\ell_1$ optimization algorithms}
\label{subse:algorithms}
The technique used in the conducted experiments replaces the $\ell_0$ pseudo-norm, found in Equation~\eqref{eqn:nphardpb}, by a ``relaxed'' $\ell_1$-norm: $\|\bm\alpha\|_1 = \sum_k | \alpha _k | $. This leads to the Basis Pursuit (BP) approach: 
\begin{equation}
\label{eqn:l1optim}
\argmin_{\bm\alpha} \|\bm\alpha\|_1
\st
\mathbf{p} = \mathbf{H D} \bm\alpha,
\end{equation}
The advantage of this proxy is that the $\ell_1$-norm is convex, allowing the use of powerful optimization algorithms. 
Furthermore, minimizing the $\ell_1$-norm under the linear equality constraint still promotes sparsity : most of the components $\alpha _k$ are pushed to zero, allowing only few non-zero coefficients. 

In practice, when noise is present, the following approach is used, that will be named L1 in the following:
\begin{equation}
\label{eqn:lasso}
 \argmin_{\bm\alpha} \|\bm\alpha\|_1 
\st
 \| \mathbf{p} - \mathbf{H D} \bm\alpha \|^2_2 \leq \epsilon.
\end{equation}
This requires tuning the data fidelity constraint $\epsilon$: 
the larger $\epsilon$, the sparser the solution $\bm \alpha$ that can be achieved, at the cost of a loss in the reconstruction accuracy. The choice of $\epsilon$ and its impact in practical situations is discussed in Section~\ref{ssse:stopcriteria}. 

   Here, we shall emphasize the parallel between Tikhonov regularization --Equation~\eqref{eqn:tikhonovregul}-- and sparse $\ell_1$ regularization --Equation~\eqref{eqn:lasso}. The latter can be expressed in Lagrangian 
   form as : 
\begin{equation}
\label{eqn:bpdn}
\argmin_{\bm\alpha} \| \mathbf{p} - \mathbf{H D} \bm\alpha \|^2_2 + \lambda \|\bm\alpha\|_1 
\end{equation}
with an appropriate choice of $\lambda$. This is known as the Basis Pursuit Denoising (BPDN) framework\cite{ChenDonohoBP}. 
Comparing Equations~\eqref{eqn:bpdn} and~\eqref{eqn:tikhonovregul}, one can see that the main difference lies in the choice of the norm: 
the $\ell_2$-norm of the Tikhonov regularization spreads the energy of the solution on all decomposition coefficients $\bm \alpha$, while the  
$\ell_1$-norm approach of BPDN promotes sparsity. 
In addition, sparse regularization gives an extra degree of freedom with the choice of dictionary $\mathbf{D}$.

While the Tikhonov solution can be expressed in closed form (Equation~\eqref{eqn:tikhonovexplicit}), the L1 method requires solving a potentially heavy optimization problem. However, numerous algorithm exist (second-order cone programming, interior point algorithms, iterative reweighted least squares, gradient projection...), and several toolboxes are available, such as SPGL1 toolbox \cite{BergFriedlander:2008,spgl1:2007} or \texttt{CVX} \cite{cvx,gb08}, used in the following.

\medskip
It should be noted that, as an alternative to $\ell_1$ optimization techniques, one can look for approximate solutions of the original problem~\eqref{eqn:nphardpb}, based on the $\ell_0$ pseudo-norm. 
This can be done using so-called greedy techniques (for instance, Orthogonal Matching Pursuit\cite{OMP}), that select atoms one by one. 

\subsection{Design of a dictionary}
\label{subse:dico}
The mechanical systems used for the validation are thin plates, point shock excited so that an acoustic impulse response field is produced (for details on  experimental setup see section~\ref{subse:settings}). They are measured in the hologram plane by a planar microphone array, (\emph{cf.} Figure~\ref{SetUp}). 

Considering a plate, the dictionary yielding the sparsest representation of the velocity field would be the collection of modal deflection shapes. Unfortunately, these can be known explicitly only in very specific cases, in particular with well controlled boundary conditions. If they were known, the problem would become considerably simpler in many aspects, but the objective of NAH imaging is often precisely to measure these modal deflection shapes.
Hence, the dictionary used here should be more generic. 

Theoretical results\cite{moiola} indicate that plane waves provide good approximations to solutions of the Helmholtz equation on any star-shaped plate (i.e. plate shapes such that there exists a point in the plate that can be connected to any point via a line segment entirely included in the plate, in particular, all convex plates are star-shaped), under any type of boundary conditions. These results have been recently extended to thin isotropic homogeneous plates \cite{cld2011}. Mathematically, the velocity of the plate $\dot{\mathbf w}$, as a solution of the Kirchhoff-Love equation, can be approximated by a sum of plane waves and evanescent waves :
\begin{equation}
\label{dec_mode1}
{\dot{\mathbf{w}}}(x,y) \approx \left(\sum_n \alpha_n e^{i \vec{k}_n \cdot \vec{x}} + 
\beta_n e^{ \vec{k}_n \cdot \vec{x}} \right)\mathbf{1}_{\cal{S}}(x,y)
\end{equation}
where $\mathbf{1}_{\cal{S}}(x,y)$ is the indicator function that restricts the plane waves to the domain $\cal{S}$ of the plate.
Here, the evanescent part is not critical for reconstruction \cite{cld2011}
and is neglected. 

\begin{figure}\center
\includegraphics[width = 8cm]{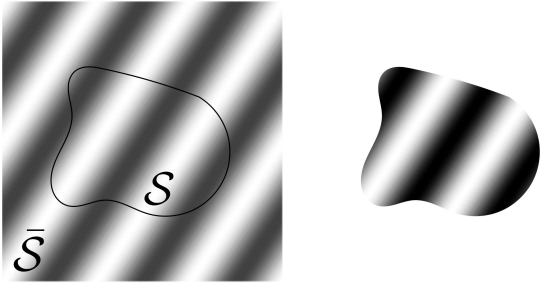}
\caption{Construction of dictionary atoms.  Left: shape $\cal{S}$ of the plate, rectangular domain $\cal{\bar S}$ used
to build the dictionary, and a Fourier basis vector on $\cal{\bar{S}}$. Right: the corresponding
atom.}
\label{atom}
\end{figure}

The construction of the dictionary $\mathbf{D}$ exploits this approximation property with plane waves. To build $\mathbf{D}$, we generate plane waves with wavevectors placed on a rectangular grid in the Fourier plane (with steps $\kappa_x$ and $\kappa_y$ in the $x$ and $y$ directions respectively, centered at the origin of the Fourier plane), and restrict them to the domain of the plate $\cal{S}$, as illustrated on Figure~\ref{atom}. This is actually equivalent to restricting to $\cal{S}$  the basis vectors of the discrete Fourier transform on a larger rectangular domain $\cal{\bar{S}}$ of size $L_x = 2\pi/\kappa_x$ and $L_y = 2\pi/\kappa_y$, containing $\cal{S}$. The main degree of freedom in the design of the dictionary is the size of $\cal{\bar{S}}$. The larger $L_x$ and $L_y$ are, the finer is the sampling of the wavevectors,
yielding  better approximations of the velocity fields, but also raising the numerical dimension of the problems to
be solved.
Unless specified otherwise, experiments will be conducted with $\cal{\bar{S}}$ twice the size of $\cal{S}$. 
Finally, as we can only solve finite size problems, we will only considered wavenumbers below a certain threshold.

\subsection{Validation of the proposed dictionary}
To validate the proposed dictionary design, we show in this preliminary experiment that typical operating deflection shapes (ODSs) can be well approximated in such a dictionary.  
By laser velocimetry, we measure 44 ODSs of a rectangular plate, between 78 Hz and 3800 Hz. 
We then compute their sparse approximation,  using only $\sharp J=8$ atoms. The approximations were found with the Orthogonal Matching Pursuit algorithm\cite{OMP}, as it allows an easier control on the sparsity $\sharp J$ than   
$\ell_1$-based optimization techniques. Note that the set of selected atoms can be different for each ODS. 
The quality of the approximation is gauged by normalized cross-correlation, as computed by Equation~\eqref{eq:correlationdef}. 

Experimental results indicate that the quality of this approximation is good: the correlation, beginning
at 99\% at 78 Hz, shows a slight decrease with frequency, but remains
always higher than
$86\%$ within the whole frequency range. 
It can be concluded that, at least in the case of a rectangular plate, a small number of atoms is sufficient to approximate the ODS, and this justifies the use of a sparse model. Further justification will be given by the experiments described in section~\ref{se:results}, first on a rectangular plate, then on a more complex D-shaped plate.

\subsection{Summary}
\label{subse:firstccl}

This section has described how the sparsity assumption with an appropriate dictionary may  replace Tikhonov regularization in the resolution of the NAH inverse problem. In section~\ref{se:results}, we will provide an experimental validation of this new framework, that shows the benefits of this alternative approach, especially in the faithful representation of high spatial frequencies.

\section{NAH and Compressive Sampling}
\label{se:csnah}

In the steps described above, we have only used an appropriate dictionary design to view the NAH regularization problem as a sparse optimization problem, but we have not changed the hardware settings nor the acquisition scheme.
However, recent theoretical advances have shown that, with some freedom on the sampling process, such sparse formulation of the problem can sometimes also help in reducing  the number of measurements needed to reconstruct a given sparse signal, sometimes well below the number of samples given by the classical Shannon-Nyquist sampling paradigm. This new theory is called Compressive Sampling\cite{CandesWakin} (CS). It can be remarked that in the literature, this theory is also called Compressed Sensing, emphasizing the parallel between sensing and sampling.

\subsection{Basics of Compressive Sampling}
\label{subse:csbasics}

Formally, compressive sampling consists of: a) {\em purposedly} reducing the number of measurements, leading to a linear system $\mathbf{p} \approx \mathbf{H} \dot{\mathbf{w}}$ with (many) more unknowns than equations; b) exploiting the sparsity of $\dot{\mathbf{w}}$ in a dictionary $\mathbf{D}$ to actually recover it through sparse regularization. 

In the general case, the under-determined linear system associated to the matrix $\mathbf{A} = \mathbf{HD}$ has an infinite number of solutions, and $\dot{\mathbf{w}}$ cannot always be estimated when only $\mathbf{p}$ is known. The key point is that, {\em under some properties} of the matrix $\mathbf{A}$, the unknown $\dot{\mathbf{w}}$ can be estimated in a stable fashion provided that it is sufficiently sparse in the dictionary $\mathbf{D}$. Interestingly, algorithms used to recover this solution are the same that are used for finding a sparse decomposition of a signal \cite{CandesWakin, BlumensathDaviesIHT, TroppGilbert}, introduced in subsection \ref{subse:algorithms}. 

\subsection{A word about theoretical guarantees}
\label{subse:randomarray}
Much theoretical work in the field of sparse regularization has shown that the so-called Restricted Isometry Property (RIP)~\cite{CandesWakin}
of the matrix $\mathbf{A}$ guarantees that sparse regularization will provide an accurate and stable estimate of $\dot{\mathbf{w}}$ from the incomplete and noisy observation $\mathbf{p} \approx \mathbf{H}\dot{\mathbf{w}}$, when $\dot{\mathbf{w}}$ is sufficiently sparse. 

The RIP condition essentially means that $\mathbf{A}$, when restricted to sparse vectors, approximately preserves the energy. Verifying that a particular matrix satisfies the RIP is considered numerically hopeless. However, many theoretical results show that various families of {\em random} matrices satisfy the RIP with very high probability. Compressed sampling exploits this property: by designing sampling systems associated to (pseudo)-random matrices, one can perform fewer measurements than the classical Shannon-Nyquist paradigm would require, while preserving the ability to reconstruct sparse data. 

In NAH, there is a priori a limited freedom in the design of the acquisition scheme. Randomness will be introduced by randomly locating the microphones in the hologram plane. Yet, the matrix $\mathbf{H}$ of the system models a propagation operator that dampens exponentially the evanescent components of the hologram. It is thus very unlikely that $\mathbf{A}$ verifies the RIP, and it is illusive to hope to fully rely on  theoretical guarantees in this setting. In practice, as we shall see, reconstruction results provided by random acquisition with a reduced number of microphones remains valid, which is in line with current knowledge in CS: practical performance remains correct much beyond theoretical guarantees (which are merely sufficient, but non necessary, conditions).

\subsection{Putting Compressed Sensing in practice for NAH?}

In the case of a signal sparse in the spatial Fourier basis, it has been shown
that few point measurements in the spatial domain are sufficient to recover exactly the signal \cite{rauhut1}, and that the reconstruction is robust to noise.
In the experimental setup considered in section~\ref{se:results}, the measurements are not strictly point measurements, but are 
more sensitive to the sources near the sensors. The theory suggests that an array with 
randomly placed sensors is a good choice of measurement scheme: in conjunction with sparse reconstruction principles, random microphone arrays perform better than regular arrays, as the measurement subspace becomes less coherent with the sparse signal subset (and therefore each measurement / microphone carries more global information about the whole experiment). 

Note that this point of view seems to contradict a recent paper \cite{BaiJSV2010}, stating that a random microphone deployment presents no particular benefit for NAH; here we must emphasize that the benefit of randomness is only apparent when employing a sparse regularization technique instead of a more standard regularization scheme.

In practice, it is difficult to manufacture a completely random array, and the precise calibration of the microphone locations on such an array can raise its share of difficulties too, as discussed further in section~\ref{se:cafeducommerce}. Hence, to assess the potential of CS for NAH, we proceed in two steps. In a first step, we use existing arrays to measure holograms on a dense regular grid, and select a random subset of microphones to simulate randomly located microphones. The corresponding detailed experiments are described in the next section. Given the success of this proof of concept, we further proceed to the design and construction of a random array. Section~\ref{se:randomarray} details the design of this array and the experimental NAH results achieved with it.

\section{Subsampling the hologram measurements}
\label{se:results}
\subsection{Setting}
\label{subse:settings}

\subsubsection{Sample structure and excitation}
An experimental validation of NAH using sparsity and CS is performed using two metallic plates. \\
{\bf Rectangular plate.} The first plate is rectangular, in aluminum, with dimensions 4mm$\times$500mm$\times$400mm. Its four corners are attached to light rubber silent blocks, so that the conditions at the boundaries approach those of a free plate.\\
{\bf D-shaped plate.} The second plate is D-shaped, made of a disk where one segment has been cut off, as shown on Figure~\ref{fig:FigODSs_circle} ; this shape 
is known to exhibit a complex modal behavior. This structure is a 4mm thick steel disk, with a 220mm radius, and the segment has been cut according to a 300mm chord. It is clamped on its side, with a clamped area of 
 20 square centimeters.

These vibrating systems are excited using an impulse hammer located on its underside, that exerts controlled point shocks (cf. Figure~\ref{SetUp}). Since the hammer is driven by an electromagnet with constant amplitude, reproducibility is ensured. The impact is located so as to excite a significant number of relevant bending vibration modes.
The acoustic impulse response measurements are carried out in a non-ideal environment, where reverberation occurs and may perturb the recordings.  

\subsubsection{Ground truth measurements}
For reference purposes, a measurement of the actual velocity field of the source is collected by a laser vibrometer, on a fine regular grid providing $50 \times 40 = 2000$ vibration impulse responses which are considered as the ground truth. For the rectangular plate, the grid has a 10 mm step along both coordinate axes. For the D-shaped plate, 1979 out of the 2000 measurements actually lie inside the plate, with steps of 8.5mm and 8.2mm in the $x$ and $y$ coordinates, respectively.

\subsubsection{Standard hologram measurements}

Hologram measurements are performed with an array of 120 electret microphones, and a custom-built 128-channel digital recorder. The standard NAH hologram is collected using a $12\times10$ regular microphone array with a 50 mm square step, the overall dimensions of the array are therefore 550 mm x 450 mm.  This basic array is moved precisely according to 16 interleaved positions in order to build a $48\times40 = 1920$ points measurement grid, with a 12.5 mm square step. The recorded signals are time-aligned.

For the rectangular plate, the array is placed at a distance $z_0 = $20mm from the plate, while for the D-shaped plate, because of the clamping system, it was located 30mm above the plate.

\subsubsection{Pre-processing}
The recorded time-domain holograms made of acoustic impulse responses are processed in order to provide holograms in the temporal Fourier domain. The resulting harmonic pressure fields or harmonic holograms can thereafter undergo the various NAH processes.
In order to ensure a better signal-to-noise ratio (SNR), we only process holograms corresponding to radiative operating deflection shapes (ODS) of the plate.  These can be found as the peaks of the temporal Fourier transform
of the impulse responses, and are scattered throughout the frequency range between 50 Hz and 4000 Hz. The complete experimental process is depicted on Figure~\ref{SetUp}.

\subsubsection{Performance measure}
The accuracy of the reconstructed velocity fields $\hat{\dot{\mathbf{w}}}$ are compared with the ground truth $\dot{\mathbf{w}}$. The performance of the various NAH techniques is evaluated using the correlation coefficient, commonly used in experimental modal analysis and defined by:
\begin{equation}
C := 
\max_{i,j}
\frac{\hat{\dot{\mathbf{w}}}^T \dot{\mathbf{w}}_{i,j}}{\|\hat{\dot{\mathbf{w}}}\|_2 \cdot \|\dot{\mathbf{w}}\|_2}
\label{eq:correlationdef}
\end{equation}
where $\dot{\mathbf{w}}_{i,j}$ is the version of $\dot{\mathbf{w}}$ shifted from $i$ pixels in the $x$-direction and $j$ pixels in the $y$-direction. This accounts for possible misfits in the location of the measurement grids between the vibrometry data and the holograms.

\subsection{Results: Tikhonov {\em vs} sparse regularization}
\label{subse:sparseregxp}
First, we propose a simple comparison between Tikhonov regularization and L1 sparse regularization, using different subsamplings of the whole set of measurements.

\begin{figure*}
\centering
    \includegraphics[width=15cm]{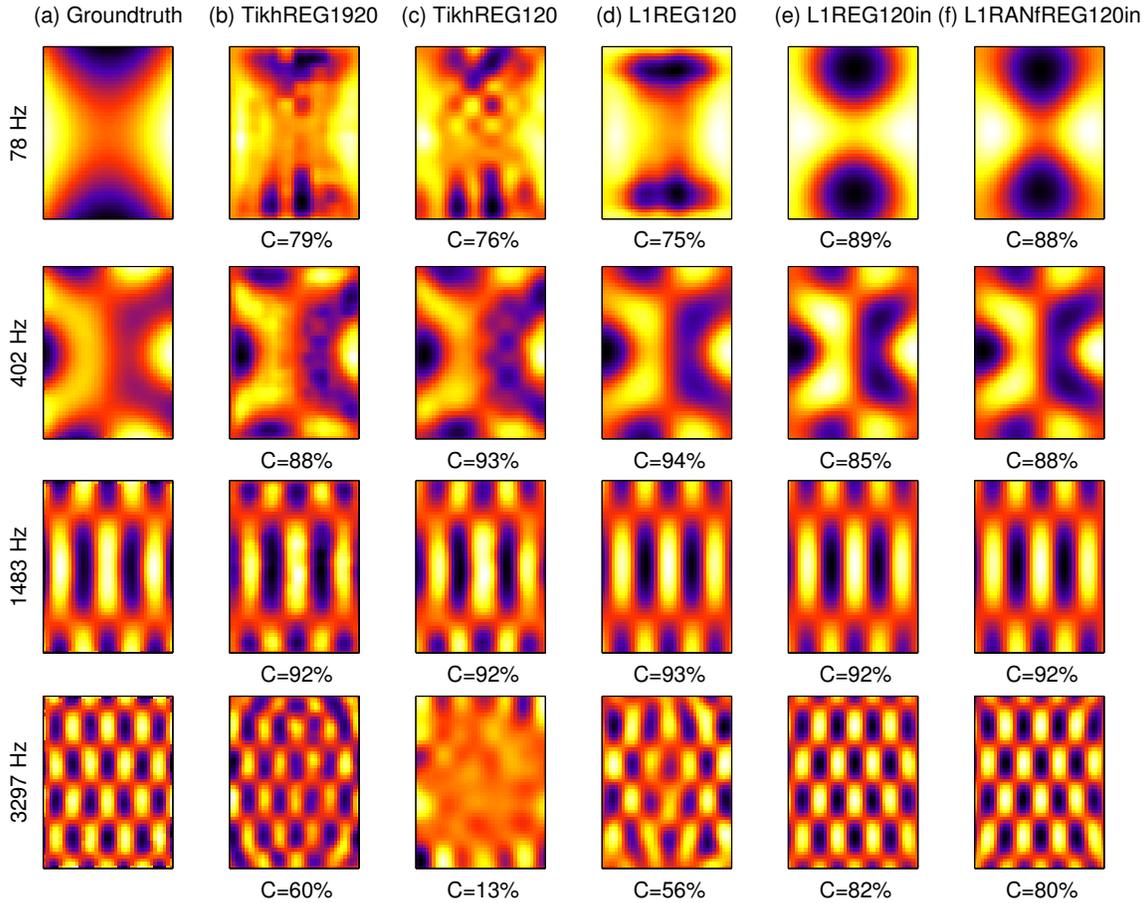} 
    \caption{Rectangular plate : Tikhonov {\em vs} L1 regularization, in various subsampling scenarii. From left to right : \protect\\
  a) Ground truth; \protect\\
    b) Tikhonov regularization + Regular array (120 microphones) at 16 positions : 1920 measurements;\protect\\
     c) Tikhonov regularization + Regular array (120 microphones) at 1 position : 120 measurements;\protect\\
     d) L1 regularization + Regular array (120 microphones) at 1 position : 120 measurements;\protect\\
     e) L1 regularization + Regular subsampling of the 1920 measurements, only ``inside'' the vibrating plate : 120 measurements;\protect\\
      f) L1 regularization + Random subsampling of the 1920 measurements, only ``inside'' the vibrating plate : 120 measurements.}
    \label{fig:FigODSs}
\end{figure*}

Figure~\ref{fig:FigODSs}-a shows four  ``ground truth'' ODSs for the rectangular plate, associated to temporal frequencies of  78 Hz, 402 Hz, 1483 Hz, and  3297 Hz. 
From the pressure hologram, we performed different types of NAH reconstruction  :
\begin{itemize}
\item Figure~\ref{fig:FigODSs}-b is obtained from the whole set of 1920 measurements (array of 120 microphones at 16 positions), using Tikhonov regularization;
\item Figure~\ref{fig:FigODSs}-c is obtained from a set of 120 measurements placed on a regular grid, corresponding to a single position of the basic array, using Tikhonov regularization;
\item Figure~\ref{fig:FigODSs}-d is obtained from a set of 120 measurements placed on a regular grid (same measurements as \ref{fig:FigODSs}-c), with L1 sparse regularization;
\item Figure~\ref{fig:FigODSs}-e is obtained from a set of 120 measurements placed on slightly tighter regular grid, so that all the measurements are above the plate, with L1 sparse regularization; 
\item Figure~\ref{fig:FigODSs}-f is obtained from a set of 120 measurements randomly selected amongst the 1920 original measurements, with all the measurements above the plate, processed with L1 sparse regularization. 
\end{itemize}

The subsampled Tikhonov approach (Figure~\ref{fig:FigODSs}-c) works well for a large range of medium frequencies, but it fails at both ends of the frequency range. At very low frequencies, the low-pass filter 
used for regularization is so narrow that its design allows leakage of higher frequencies (a drawback that does not depend on the number of measurements, as seen on Figure~\ref{fig:FigODSs}-b). At very high frequencies, 
the spatial sampling of the pressure hologram does not satisfy the Shannon-Nyquist criterion, and therefore spatial aliasing occurs.  
On the opposite, NAH results obtained with L1 sparse regularization (Figure~\ref{fig:FigODSs}-d and Figure~\ref{fig:FigODSs}-e) reproduce faithfully the ODSs with the same number of microphones, 
even at high spatial frequencies. 
Interestingly, in the case of L1 sparse regularization it is better to concentrate the measurements strictly ``inside'' the plate (Figure~\ref{fig:FigODSs}-e).
This is explained by the fact that, although the plate is unbaffled, the propagation model assumes a baffled plate. The mismatch between these two models
is acceptable for the microphones directly above the plate, but is much larger for the microphones outside the plate aperture. Therefore, discarding
these measurements yields better reconstructions. Moreover, keeping the same number of microphones on a smaller area allows for a slightly denser grid. 
The last column (Figure~\ref{fig:FigODSs}-f) shows that similar results are obtained when the measurements are randomly selected amongst the 1920 original measurements (ensuring that all the selected 
measurements are above the plate). 
 
\begin{figure*}[htbp]
\centering
    \includegraphics[width=15cm]{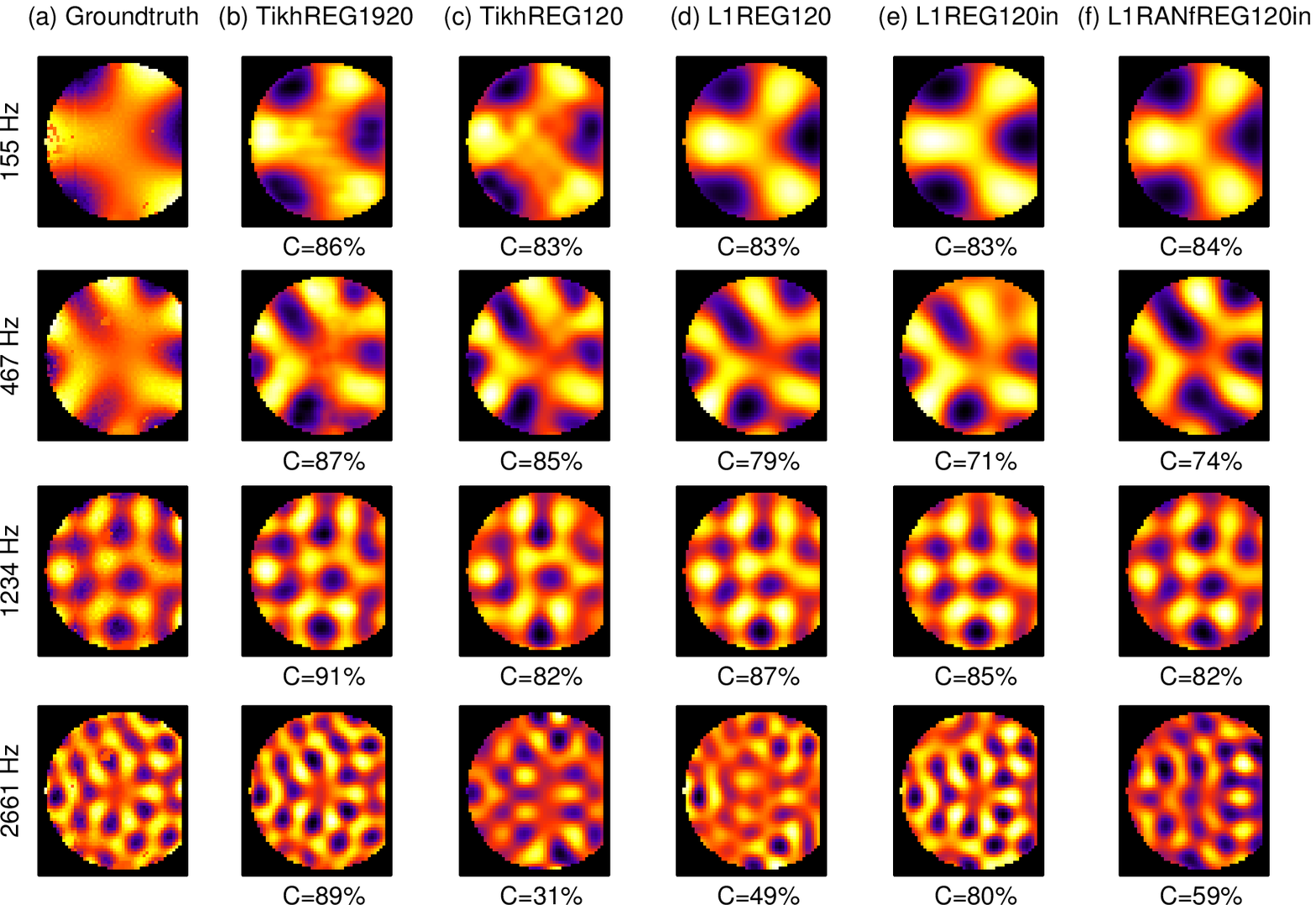} 
    \caption{D-shaped plate :  Tikhonov {\em vs} L1 sparse regularization. Same subsampling as on Figure~\ref{fig:FigODSs}.}
    \label{fig:FigODSs_circle}
\end{figure*}

\medskip
To validate the technique in a more general case, we used the same experimental protocol for the D-shaped plate. Results are shown on Figure~\ref{fig:FigODSs_circle}. 
Although the shape and the boundary conditions in these case are more complex, the exact same conclusions can be drawn, at least qualitatively. However, the associated ODSs are less sparse, and therefore slightly more measurements 
are needed in order to recover them accurately.

\subsection{Results: regular {\em vs} random array}

\begin{figure}[htbp]
\centering
 \includegraphics[width=8.5cm]{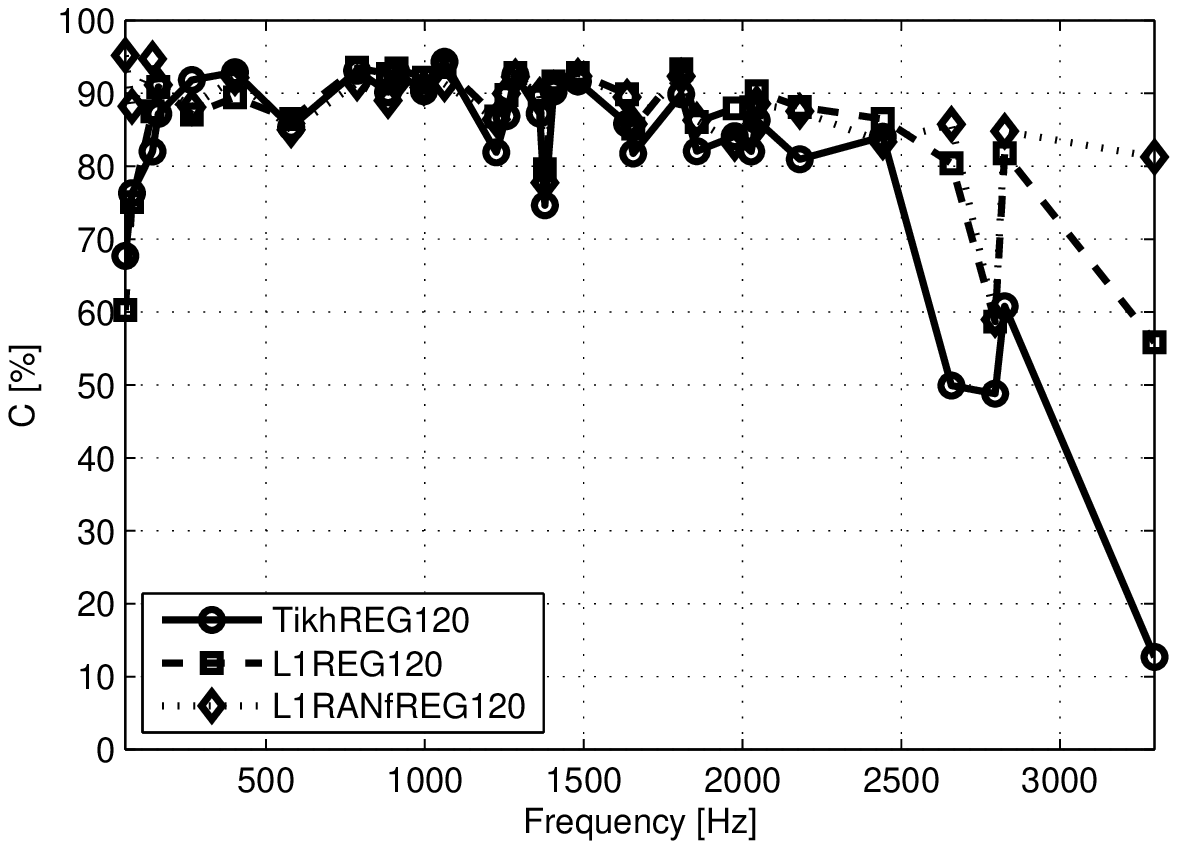}
  \caption{Comparison of 3 approaches for the NAH identification of the radiating ODS up to 3.3kHz  : Tikhonov regularization with regular array (120 measurements), L1 sparse  regularization with regular array  (120 regularly spaced measurements ``inside'' the plate), and L1 sparse regularization with random subsampling  (120 measurements randomly selected from a regular grid 1920 measurements, restricted to be over the plate). }
     \label{CompStdCSPlot}
\end{figure}
In order to validate these results in a more quantitative manner,  we compare on the whole frequency range the different spatial subsampling strategies (random or regular),  for the two regularization techniques (Tikhonov or L1 sparse). 
Figure~\ref{CompStdCSPlot} compares  standard NAH using the regularly subsampled array (120 measurements),  L1 sparse regularization using a regular array of 120 measurements above the plate taken from the full set of 1920 NAH measurement, and L1 sparse regularization using 120 measurements above the plate randomly drawn from the whole set of 1920 measurements. 
Whereas all the algorithms have similar performance for medium frequencies, the accuracy of Tikhonov-based NAH drops in the low and high frequency ranges.  L1 sparse regularization improves significantly on both sides, with slightly better performance for the random array. 

\section{Use of a random array}
\label{se:randomarray}

 The previous section has shown that it is possible to maintain good reconstruction 
 results while randomly undersampling the pressure hologram measurements. In this section, we put this idea into practice by constructing an array of randomly located microphones. 
  The main advantage of using this array is that we can perform all the measurements at once, with a single excitation, as opposed to the regular array that had to be moved in 16 interleaved positions in order to finely grid the hologram plane, and hence required a series of 16 impulse excitations and measurements. 
   However, uniformly distributed random arrays are difficult to build for practical reasons (microphone mounts), and therefore we performed numerical simulations for the design of 
   simpler random arrays. These arrays have the advantage that they can be built using several straight bars. 
\subsection{Considered array designs}
\begin{figure}[htbp]
\centering
\includegraphics[width=7cm]{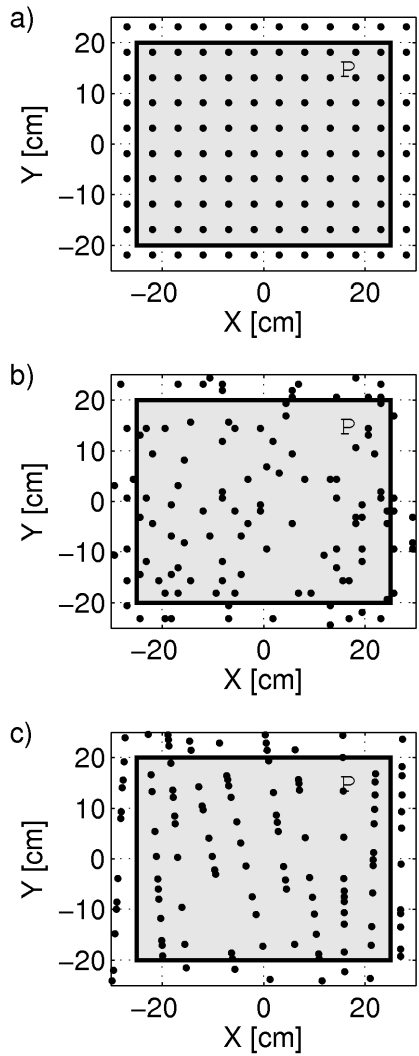}
\caption{Examples of microphone distribution for three design strategies: a) regular array, b) random array c) pseudo-random oblique array. The area colored in gray is the area of the hologram plane located exactly at the vertical of the vibrating plate. The dots indicate the location of the microphones.}
 \label{AntennaPlot}
\end{figure}

The considered array designs are qualified according to their arrangement: regular, tensorial, parallel, crossed oblique, oblique, random from regular, or random, and are defined as follows:
\begin{itemize}
  \item regular: a regular array, that is computed for reference. In practice, for example, 120 microphones would be placed on $I=12$ bars parallel to the $y$-axis; each bar would hold $J=10$ microphones evenly spread.
  \item tensorial: microphones are located at all positions $(x_i, y_j)$, where $\{x_i\}_{i=1\dots I}$ and $\{y_j\}_{j=1\dots J}$ are uniformly drawn in the $(0,1)$ interval. The $x_i$ determine the positions of the bars, and the  $y_j$ the position of the microphones along the bars. 
  \item parallel: microphones are located at all positions $(x_i, y_{j,i})$, where $\{x_i\}_{i=1\dots I}$ and, for every $i$,  $\{y_{j,i}\}_{j=1\dots J}$ are uniformly drawn in the $(0,1)$ interval. Here, the position of the microphones is different along each bar.  
   \item  crossed oblique, same as parallel, but the bars are not constrained to be parallel to the $y$-axis.
    \item  oblique, same as crossed oblique, but the bars are constrained not to cross each other.
    \item random from regular: a random selection of microphones located of a denser regular array. 
\item random : purely random location of the microphones, at positions $\{ (x_k, y_k) \}_{k=1\dots I \times J}$,  uniformly drawn in $(0,1)\times(0,1)$. Note that a practical issue using this array would be to calibrate the exact location of each microphone. 
\end{itemize}
Figure~\ref{AntennaPlot} shows examples of such arrays in the regular, random, and oblique arrangements. 

\subsection{Numerical simulations}
\label{NumSim}
Reconstructions have been simulated for these test arrays, with varying numbers of sensors. The simulated plate is a simply supported square steel plate, with dimensions 1mm$\times$200mm$\times$200mm. The simulated hologram and the plate are separated by 20mm. The domain used to build the dictionary has dimensions twice the dimensions
of the plate. 
It should be noted that the modes
of the plates are not exactly sparse in the dictionary used here.

\begin{figure}[htbp]\center
\includegraphics[width = 7cm]{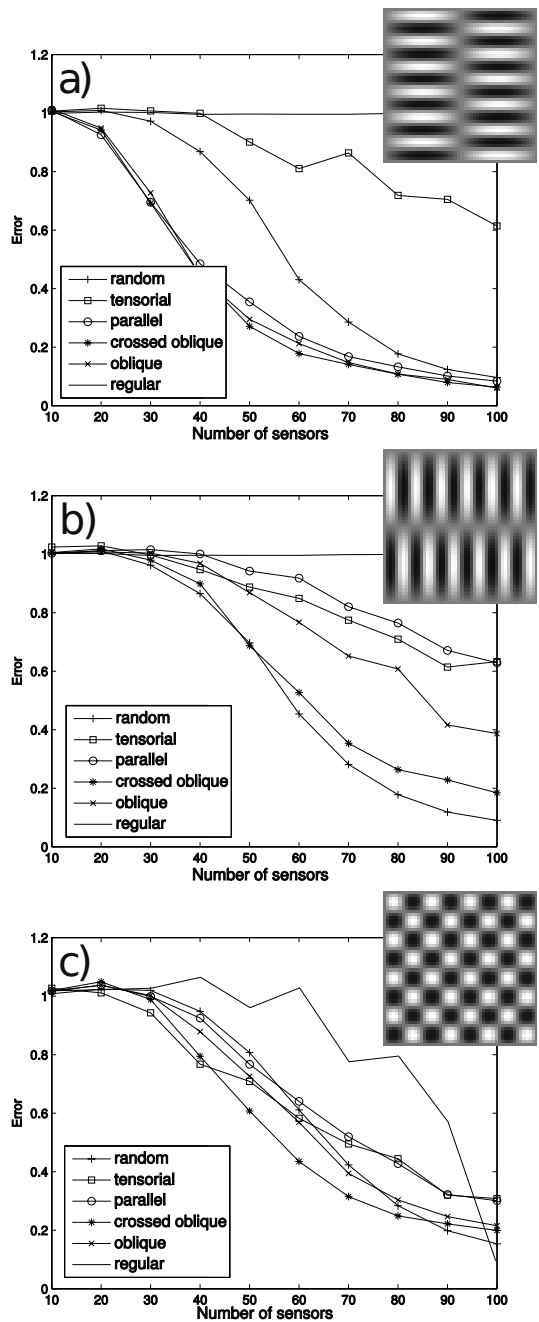}

\caption{Average reconstruction error with respect to the number of microphones, for different types of arrays. a) (2,12)-mode at 1053 Hz. b) (12,2)-mode at 1053 Hz. c) (8,8)-mode at 911 Hz.}
\label{simu2812}
\end{figure}

Figure~\ref{simu2812} shows the reconstruction error (as defined in Equation~\eqref{eq:correlationdef})
obtained with arrays of 10 to 100 sensors, for each type of considered array, for the modes (2,12) and (12,2), at 1053 Hz, and (8,8) at 911 Hz. For each type of random array and each number of sensors, the error is averaged over 30 different draws of the random array. Reconstructions are obtained by $\ell_1$ optimization as described above, solving the Basis Pursuit problem~\eqref{eqn:l1optim}
with the \texttt{CVX} \cite{cvx,gb08} package.

The regular arrays fail to yield a correct reconstruction
of the (2,12) and (12,2) modes, because the spatial Shannon-Nyquist criterion
is not met with 100 or less measurements. Results with the regular arrays are better
for the (8,8) mode as the Shannon-Nyquist criterion is met in this case. 

The uniformly distributed  random arrays have good performance even for sub-Nyquist measurements,
but, as pointed before, are cumbersome to build. 

Other arrays have varying performances, the best on average being here the
crossed oblique array. 
It should be noted that the performance of the
 parallel, crossed oblique and oblique arrays are different for the (2,12) and (12,2) modes which differ only by their orientation. 
It is not unexpected, since these array are not invariant by a 90\textsuperscript{o}-rotation.

Finally, the {\em oblique} array (Figure~\ref{AntennaPlot}-c) leads to a good tradeoff  between reconstruction fidelity and hardware feasibility. Therefore, this design was chosen for the experimental validation, which is presented next.

\subsection{Experimental validation}

\begin{figure*}[htbp]\center
\includegraphics[width=10cm]{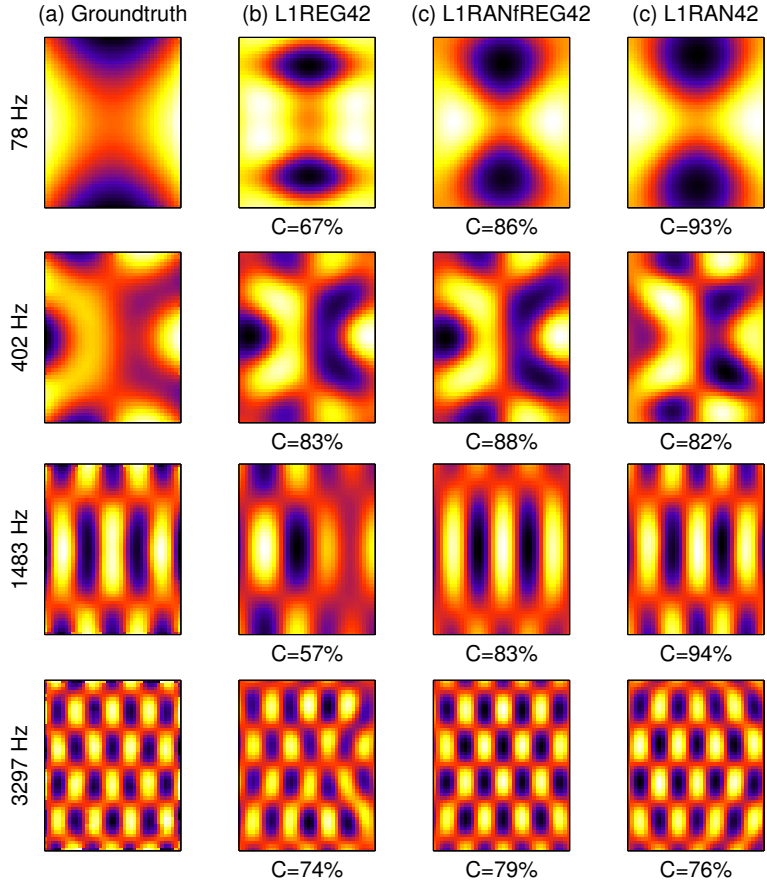}

\caption{Comparison of spatial subsampling strategies with L1 sparse regularization. From left to right : \protect\\
  a) Ground truth; \protect\\
    b) Regular array (42 microphones) over the plate;\protect\\
     c) Random selection of 42 microphones from regular set of 1920 measurements over the plate;\protect\\
     d) 42 microphones from oblique array.}
\label{fig:randomarray}
\end{figure*}

To evaluate the impact of the array design on the reconstruction accuracy, we compared on the whole frequency range three subsampling strategies, using L1 sparse regularization for reconstruction. The results are displayed on Figure~\ref{fig:randomarray}. Figure~\ref{fig:randomarray}-a shows the ``ground truth'' ODSs acquired with the experimental setting described in section~\ref{se:results}. The reconstructions were obtained from measurements performed {\em over the plate} as follows:
\begin{itemize}
\item Figure~\ref{fig:randomarray}-b is obtained from 42 measurements placed on a regular grid, corresponding to a regular subsampling of the 1920 NAH measurements;
\item Figure~\ref{fig:randomarray}-c is obtained from 42 measurements randomly selected from those of the 1920 microphones that were above the plate;
\item Figure~\ref{fig:randomarray}-d is obtained from 42 measurements selected from the 80 microphones of the implemented oblique array that were above the plate.
\end{itemize}
We obtained similar results when varying the number of microphones above a minimum number of microphones that depends on the reconstructed ODS.

\subsection{Tuning the Compressive Sensing reconstruction}

\subsubsection{Choice of the algorithm}
As an alternative to the L1 approach used so far, one can use other classes of sparse solvers. A popular choice is the Orthogonal Matching Pursuit\cite{OMP} (OMP), that builds an approximate solution of the original sparse problem 
\eqref{eqn:nphardpb} in an iterative manner : a first atom is selected, its contribution is removed, and the process is iterated on the residual until a number $J$ of atoms have been selected. Besides its simplicity, a potential advantage of OMP is that the sparsity level can be freely chosen. 

In practice, the choice of a sparse regularization algorithm to perform NAH reconstruction depends on several factors. 
As we have seen, sparse regularization relies on the design of a dictionary in which the source is sufficiently sparse. Moreover, any regularization technique requires the tuning of a regularization parameter, which is dependent both on the source under study and on the algorithm. The robustness of the reconstruction results when these parameters are ``blindly'' tuned is therefore an important criterion to drive the choice of an algorithm.

\subsubsection{Tuning of the dictionary}

\begin{figure}[htbp]
		\centering
    \includegraphics[width=8.5cm]{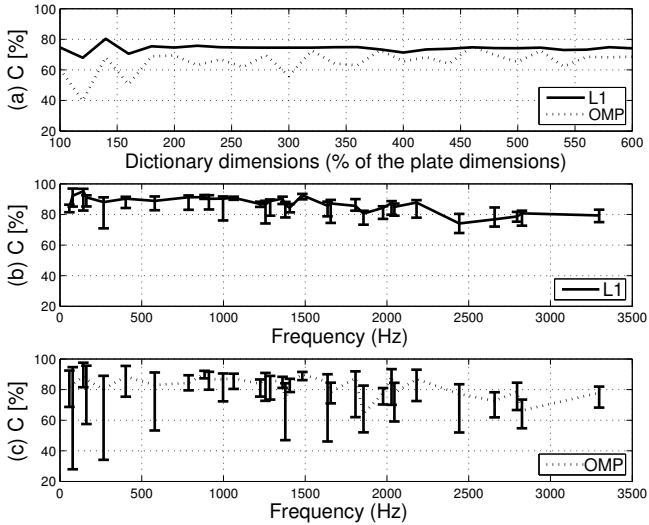}
    \caption{Comparison of the performance of the reconstruction algorithms L1 an OMP in the identification of the radiating ODS. a) effect of the dictionary dimensions; b) variations with ODS frequency for L1; c) variations with ODS frequency for OMP.}
    \label{CompBPandOMPPlot}
\end{figure}

For the identification of the considered free plate's ODSs, as said before, the dictionary consists in the spatial Fourier atoms of a surface $\cal{\bar{S}}$ larger than the plate. When varying the dimensions of this $\cal{\bar{S}}$ between $100\%$ and $600\%$ of that of the plate we found that the correlation with the reference ODS was very stable with L1 reconstruction. When using OMP with a fixed stopping criterion, the correlation varied more significantly with the size of $\cal{\bar{S}}$. This is illustrated on Figure~\ref{CompBPandOMPPlot}-a.

Figure~\ref{CompBPandOMPPlot}-b (resp. Figure~\ref{CompBPandOMPPlot}-c)  shows the typical reconstruction accuracy as a function of the frequency, with L1 (resp. with OMP). The error bars show the extremal values of the reconstruction accuracy for each frequency when the dimensions of $\cal{\bar{S}}$ are varied between $100\%$ and $600\%$ of those of the plate.
This corresponds to NAH reconstruction with sparse regularization from recordings made with about 80 microphones of the random array of section~\ref{se:randomarray}. 

One can observe that the L1 algorithm behaves stably whatever the spatial extent of the dictionary, while the performance of OMP depends heavily on the fine tuning of the dictionary, and is hard to control a priori.  We conclude that, even if OMP is often more efficient in terms of computation time, L1 should be prefered for its better robustness with respect to the tuning of the dictionary.

\subsubsection{Tuning of the regularization parameter}
\label{ssse:stopcriteria}

\begin{figure}[htbp]
\centering
    \includegraphics[width=8cm]{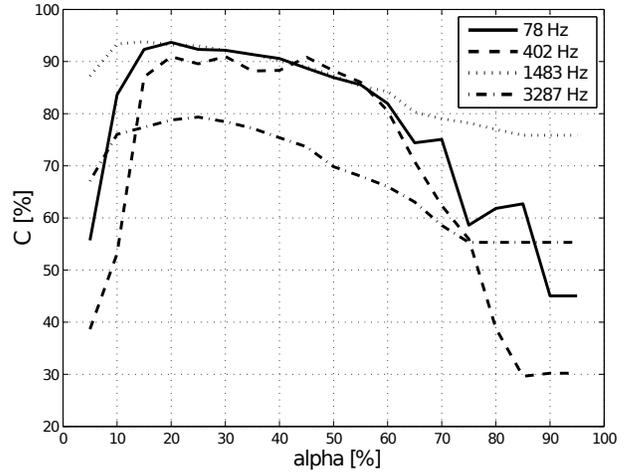}
    \caption{Variation of correlation coefficient according to the input noise parameter $\epsilon = \alpha  \|p\|_2$}
    \label{epsilon}
\end{figure}

Because of the above highlighted lack of robustness of OMP with respect to the tuning of the dictionary, we now focus on L1.
When using the L1 algorithm, the input error parameter $\epsilon$ is required, as described in section~\ref{subse:algorithms}. This parameter is defined as a function of the measurement noise (acoustic background, electronic noise), the microphones positioning error, the propagation model error and the approximate sparsity of the source. Since these items are not easily quantifiable, except for the background and electronic noises, it is not obvious to set objectively the value of $\epsilon$ so that the algorithm stops when the most accurate approximation of the source is found.
Nevertheless, by studying the correlation between reconstructed velocity and the ground truth (see Figure~\ref{epsilon}), we can see that a relatively large range of $\epsilon$ is acceptable. $\epsilon$ is expressed as a function of the hologram norm: $\epsilon = \alpha  \|p\|_2$. It appears that a level of 20\% to 30\% of the signal norm is acceptable for the 4 selected frequencies.
However, an objective method for the evaluation of $\epsilon$ is still to be designed to get free from the experimental conditions.

\section{DISCUSSION}
\label{se:cafeducommerce} 

\subsection{Improvements of NAH by CS}
This subsection emphasizes several improvements of NAH brought by the use of compressive sampling and proposes some interpretations of previously observed results.  

\subsubsection{Calibration errors}
At low frequencies, the Tikhonov-based NAH reconstruction exhibits large errors in the estimation of the amplitude (see Figure~\ref{fig:FigODSs}-b/c) at 78 Hz and 402 Hz).  This may come from the straightforward dependence of reconstructed ODSs on the SNR of the measured acoustic pressures. Indeed, pressure measurements suffer from background and electronic noise, but also from possible calibration errors.

Sparsely regularized NAH is less dependent on the sensors amplitude errors, since the identified ODSs are linear combinations of atoms picked up in a deterministic dictionary. Of course, the algorithm will not succeed in catching the relevant atoms if the amplitude errors are too strong, especially at high frequencies. However, when only a few microphones are affected by small calibration errors, the reconstruction is not badly corrupted, as illustrated on Figure~\ref{fig:FigODSs}-d/e/f.

\subsubsection{Hologram extent}
The experimental standard NAH process is performed over a finite measurement aperture. This rises high wavenumbers components representing the edges of the aperture. Backpropagation exponentially amplifies these components, distorting the reconstructed source. In order to lower these effects, the acoustic pressure field is recorded over a surface larger than the source, so as to reach null pressure points. The edges of the obtained hologram are thereafter smoothed using a flat top tapering window. However, such an extension of the hologram implies more microphones or additional measurements.

The above NAH experiments using sparse regularization show that the reconstruction is poorer when the microphones outside the vibrating domain are considered. Therefore, these measurements are discarded and only the microphones located above the source are used.
One possible explanation for this result lies in the fact that the propagation model considers implicitly a baffled plate,
while in the experiments, the plate was unbaffled. The mismatch between the two models being larger for the microphones
outside the plate aperture, better results are obtained by discarding these measurements.

\subsubsection{Robust regularization}
The plates under study have free edges, where large amplitudes of vibration occur. This leads to sharp discontinuities of the source vibration field, whose signature in the $k$-space is located at high spatial wavenumbers.  With the improved Tikhonov regularization using the low-pass filter described in section~\ref{sse:StdRegul}, this high-$k$ area is filtered out, which artificially smoothes the discontinuities. This is more critical for the low frequency ODSs where the evanescent components are prominent and for which the cut-off wavenumber is lower. This issue is illustrated on Figure~\ref{fig:FigODSs}-b/c especially at 78 Hz.

Eventually, sparse regularization strategies should improve the reconstruction of discontinuities, since they do not use any low-pass filtering in the $k$-space. Moreover the atoms of the dictionary present discontinuities at there edges which should contribute to a better identification of the free edges. This is not observed systematically but is obvious on Figures~\ref{fig:FigODSs}-e/f. 

\subsubsection{Spatial sampling and aliasing}
For standard NAH measurements, the pressure field is sampled according to a regular mesh with whose step sets the $k-$space limit of occurrence for spatial aliasing. For the rectangular plate, the ODSs hold vibrating components with wave numbers higher than the limit for frequencies above 2450Hz. This explains that the regular array combined with the Tikhonov regularization fails at higher frequencies (cf. Figure~\ref{CompStdCSPlot} - TikhREG120).

Interestingly, sparse regularization with regular sampling provides correct reconstructions. 
Here again the deterministic nature of the dictionary atoms avoids aliasing. Nevertheless when the acoustic field is highly undersampled using a regular array, the lack of information does not allow a correct reconstruction (see Figure~\ref{fig:randomarray}-b))

\begin{figure}[htbp]
\centering
 \includegraphics[width=8cm]{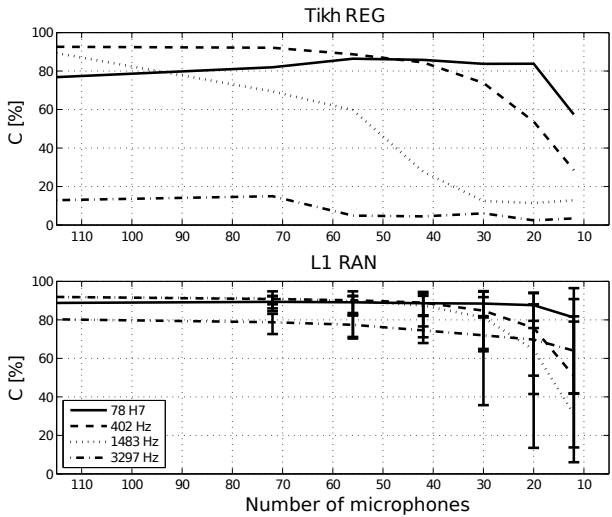}
 \caption{Accuracy of ODSs identification at different frequencies  with respect to the number of microphones. Top: Regular array with Tikhonov regularization; Bottom: Random array with L1 sparse regularization over 100 averaged draws.  }
 \label{SamplingLimit}
\end{figure}

An advantage of random sampling in the sparse regularization case is that it 
always allows a significant downsampling of the acoustic field. In order to highlight this fact, the number of microphones used for the CS process is decreased until the reconstruction fails. Figure~\ref{SamplingLimit} shows the evolution of the correlation coefficient in the low, mid and high frequency range, with respect to the number of microphones for two competing NAH methods, with the rectangular plate. Considering the random array using L1 sparse regularization (Fig.~\ref{SamplingLimit}-b)), for each number of sensors, 100 random draws of the set of microphones are performed out of the pseudo-random array and the 100 subsequent correlation coefficients are averaged. The average remains quasi constant down to about 40 microphones, regardless of the frequency on all the band of interest (see also Figure~\ref{fig:FigODSs}-f and Figure~\ref{fig:randomarray}). Thus, using about 40 measurements is enough to identify the ODS at all frequencies with a very good accuracy (C over 80\% in average), while the regular array using Tikhonov regularization fails according to the Nyquist limit dictated by the number of sensors (Fig.~\ref{SamplingLimit}-a)). Note that with 40 randomly placed microphones using sparse regularization, in the high frequency range (Figure~\ref{fig:randomarray}) at 3297 Hz), the spatial sampling is much lower than the Nyquist limit, yet the sources are accurately identified.

\subsubsection {Algorithms complexity}
Comparing the methods used in this work brings a few remarks regarding the complexity of the algorithms, in terms of computation speed. When the Tikhonov regularization parameter is known, at a given frequency, Standard NAH is by far the fastest process. %(~0.01s)
However, an exhaustive search step of this parameter is possibly needed  at each frequency of interest, which severely slows down the computation.  As for CS using the $\ell_1$ minimization process (L1) for the reconstruction, it is significantly more demanding in terms of computational cost, but typical computation time remain reasonable (of the order of 1s per ODS, on a standard PC).

\section{CONCLUSIONS}
\label{se:conclu} 

This paper presents some experiments that illustrate the use of sparsity principles for NAH, that are valid for any star-shaped homogeneous plate.  As a first 
contribution, sparsity principles have been shown to act as an alternative regularization technique. Its main benefit is that it can easily be tailored 
to specific geometries and boundary conditions, while not suffering from the typical spatial low-pass filtering effect - this is particularly useful at low frequencies. Secondly, this can be used to obtain good results with generally a lower number of microphones than standard NAH, using the recent paradigm of compressive sensing: using randomly placed microphones, NAH can still be performed at frequencies above the spatial Nyquist rate. 
The main difficulty in using  sparsity-promoting techniques is that recovering the desired Operational Deflection Shapes can be significantly more demanding than in the case of standard NAH, as it requires solving a large optimization problem under sparsity constraints. However, a number of such algorithms have been developed in the last few years, and can usually be used without substantial modification.  

\medskip

While the method has been demonstrated with homogeneous,  isotropic and convex plates, it can be extended to a wider class of structures.
Vibrations of anisotropic plates are also sparse in the dictionary defined in section \ref{subse:sparsity}. Vibrations of structures with cylindrical or spherical shapes can be described by a similar model, with 
measurements taken over a conformal surface around the source. Plates with inhomogeneous thickness can also be handled numerically by the Vekua theory.  
Finally, for plates with complex  geometries (possibly non-convex), that can be divided into simple star-shaped subdomains, a similar approach can be pursued, provided that continuity constraints are incorporated into the inversion.
The experimental validation of these extensions is left for further research.
Another extension of the method would be, for homogeneous plates, the application
of a more constrained model, where the plane waves used in the decomposition
of the velocity field are constrained to have wavevectors of the same
magnitude\cite{cld2011}.

   As a concluding remark, the Compressive Sensing framework is a good illustration of an increasing trend in imaging techniques, to jointly optimize the sensing procedure together with the data processing algorithms. In this respect, this contribution gives a new insight about the somehow controversial issue whether NAH arrays should be random or not \cite{BaiJSV2010}: in our setting, randomness is useful, but only when combined with a sparse regularization technique.

%% BEGIN_FOOTER %%

\section*{Reproducible research}
To comply with the reproducible research principles\cite{ReprodRes}, the datasets and code used in this article can be found at the following URL : {\tt http://echange.inria.fr/nah/}

%\newpage

\section*{Acknowledgments}
This work was supported by the Agence Nationale de la Recherche (ANR),  ECHANGE project (ANR-08-EMER-006). 

\bibliographystyle{jasanum}

%\newpage

%% BIBLIO %%

\bibliography{biblio}
\end{document}